\documentclass[fleqn,a4paper,11pt]{article}
 
\usepackage{color}
 
\usepackage{graphicx}
\usepackage{amssymb}
\usepackage{lineno}
\usepackage{wrapfig}
\newcommand \mumu {$\mu^+\mu^-$ }

\newcommand \pt {$p_T$ }

\newcommand \sqn {$\sqrt{s_{_{NN}}}$ }
 
\newcommand \dau {d+Au }
\newcommand \auau {Au+Au }
\newcommand \cucu {Cu+Cu }
\newcommand \ppb {p+Pb }
\newcommand \pbpb {Pb+Pb }
\newcommand \npart {$N_{part}$ }

\newcommand \sqnr {$\sqrt{s_{_{NN}}}$~=~200~GeV }

\newcommand \raa {$R_{AA}$ }
\newcommand \jpsi {J/$\psi$ } 

\def\JPG{{J. Phys}~{\bf G}}

\def\NPA{{Nucl. Phys.}~{\bf A}}

\def\PLB{{Phys. Lett.}~{\bf B}}

\def\PRL{Phys. Rev. Lett.\ }

\def\PRC{{Phys. Rev.}~{\bf C}}

\def\EPJC{{Eur.~Phys.~J.}~{\bf C}}

\def\JHEP{JHEP }

\begin{document}
\title{\bf {The quarkonium saga in heavy ion collisions}}
 
 \author{Itzhak Tserruya\footnote{email: Itzhak.Tserruya@weizmann.ac.il} \\   
                    Weizmann Institute of Science, Rehovot, Israel} 
 
\maketitle 
    
\begin{abstract}
\medskip
\jpsi suppression was proposed more than 25 years ago as an unambiguous signature for the formation of the Quark Gluon Plasma in relativistic heavy ion collisions.
After intensive efforts, both experimental and theoretical, the quarkonium saga remains exciting, producing surprising results and not fully understood. 
This talk focuses on recent results on quarkonium production at RHIC and the LHC.
\end{abstract}

\section{Introduction} 
\label{introduction}
More than 25 years ago, Matsui and Satz published their by now classic paper 
where they proposed \jpsi suppression as an unambiguous signature of quark deconfinement in the Quark Gluon Plasma (QGP) \cite{matsui-satz}.
At high color density the confining potential becomes color screened, 
(the QCD equivalent of the QED Debye screening of electrical charges) 
effectively limiting the range of the strong interaction. As a consequence, when the screening radius becomes smaller than the $c\overline{c}$  binding radius,
the $c$ and $\overline{c}$ cannot   bind together any longer,
leading to suppression of the \jpsi yield in nuclear collisions. 
 
One of the first observations made by the NA38/NA50 experiment in the framework of the CERN SPS heavy ion program,  
was indeed the suppression of \jpsi production in S+U collisions at 200 A GeV \cite{na38-SU}. However, this suppression was found to follow the same systematic trend observed in proton-nucleus collisions and in collisions involving light nuclei like O+Cu and O+U. The \jpsi suppression in all these systems was properly accounted for by a final state absorption cross section of $\sigma_{abs} \sim$ 4 mb  of the charmonium state in nuclear matter \cite{na50-pA}. 
An anomalous \jpsi suppression, stronger than expected from this absorption cross section,
was observed in semi-central and central Pb+Pb collisions at 158 A GeV
suggesting the onset of an additional suppression mechanism, possibly the production of a deconfined state of matter \cite{na50-PbPb}. Consistent results were later obtained by the NA60 experiment in In+In collisions at 158 A GeV \cite{na60-jpsi}.

Experimental results from RHIC and more recently from LHC, unveiled a much richer physics landscape with a variety of competing effects that can potentially affect the charmonium production in nuclear collisions. These include nuclear modifications of the gluon distribution functions (shadowing or anti-shadowing), gluon saturation, initial and final state $k_T$ scattering, initial and final state parton energy loss, 
nuclear absorption, co-mover breakup and recombination.
 
This paper focuses on some of the most recents quarkonium results obtained  at RHIC and the LHC.

\section{RHIC results}
\label{rhic}
\begin{figure}[h!]
     \begin{center}
           \includegraphics*[width=70mm] {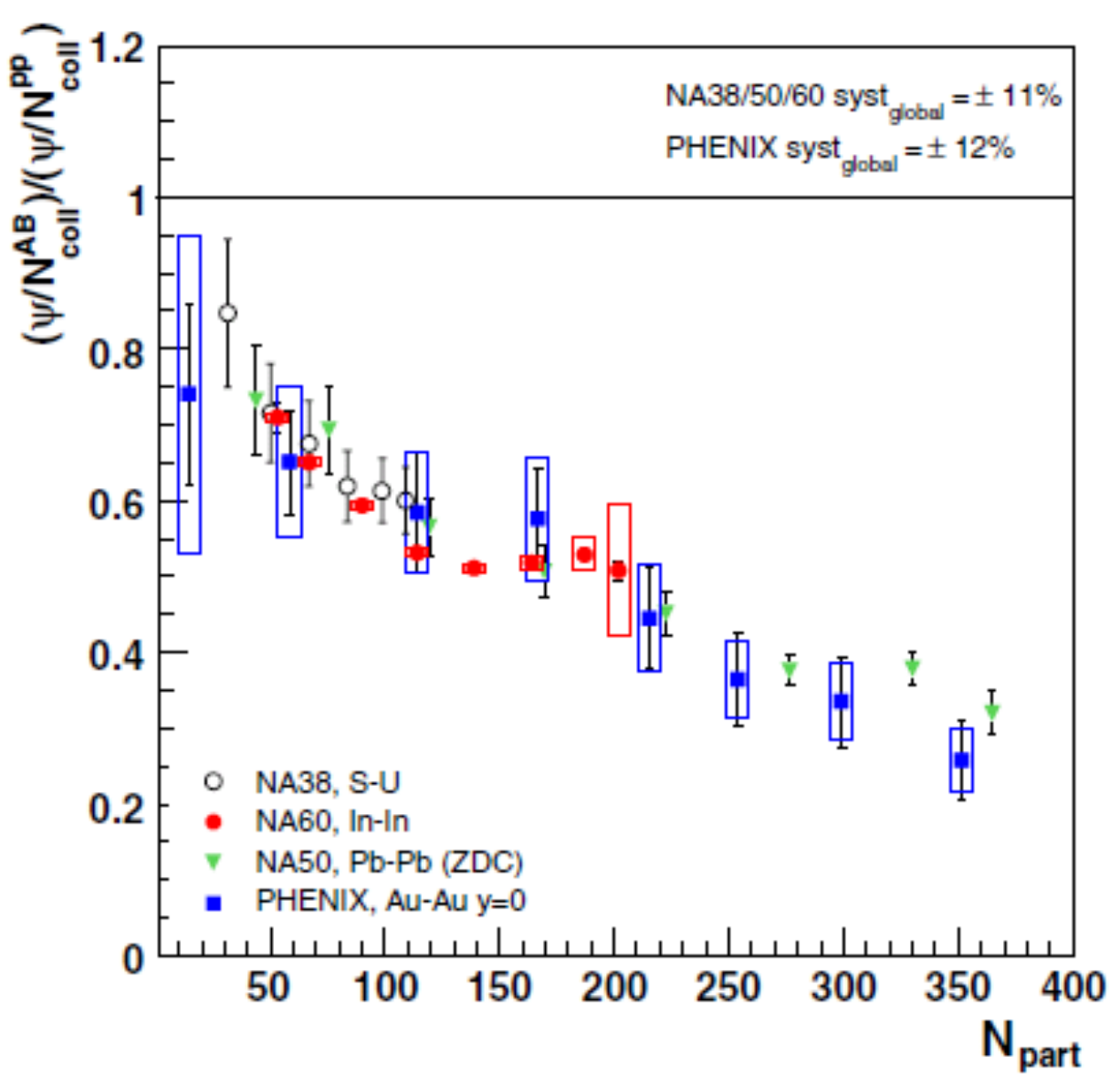}             
           \includegraphics*[width=70mm, height=65mm] {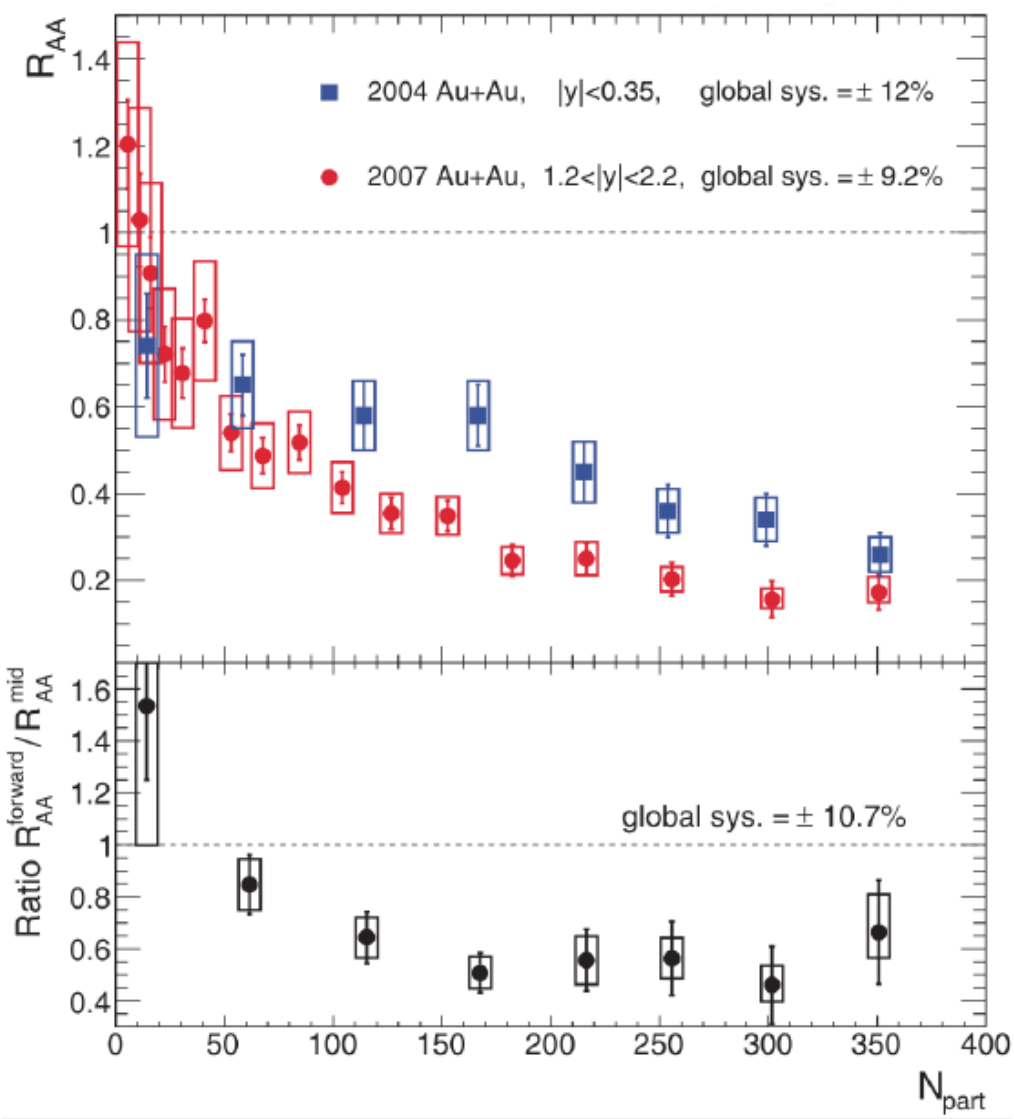}
\vspace{-2mm}
\caption{Left: The \jpsi nuclear modification factor, R$_{AA}$, measured at mid-rapidity by PHENIX at \sqnr \cite{phenix-jpsi-200} and by NA38, NA50 and NA60 at SPS energies \cite{jpsi-sps}.  Right: \jpsi \raa measured by PHENIX at forward and mid-rapidity at \sqnr \cite{phenix-jpsi-200}.}
           \label{fig:rhic}
     \end{center}
\vspace{-4mm}
\end{figure}

The first charmonia measurements performed by the PHENIX experiment at RHIC in \auau collisions at \sqnr  yielded two surprising results. First, at mid-rapidity, the level of \jpsi suppression, quantified by the the nuclear modification factor \raa (defined as the ratio of the yield per binary nucleon-nucleon collision, $N_{coll}$, in A+A collisions to the yield in p+p collisions), is very similar to the one observed at the SPS (see left panel of Fig.~\ref{fig:rhic}) \cite{phenix-jpsi-200, jpsi-sps}. This is contrary to the stronger suppression anticipated at RHIC due to the increase of more than one order of magnitude in collision energy.
The second surprising result is that \jpsi is more suppressed at forward than at mid-rapidity by about a factor of two, in spite of the fact that the energy density at forward rapidity is presumably smaller than at mid-rapidity (see right panel of Fig.~\ref{fig:rhic}) \cite{phenix-jpsi-200}.
 
\begin{figure}
\begin{center}
            \includegraphics*[width=90mm] {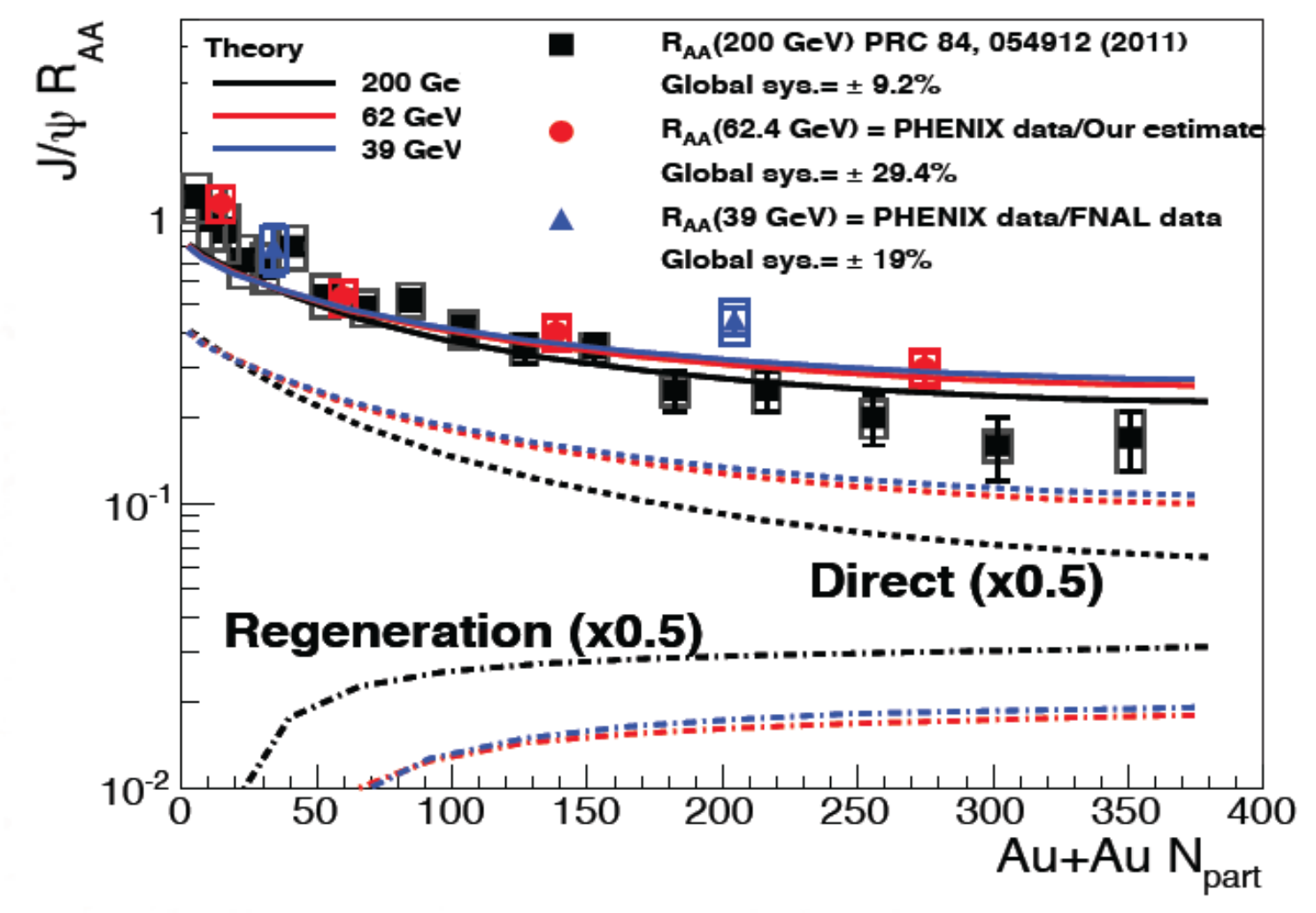}
            \caption{\jpsi \raa measured by PHENIX at mid-rapidity at \sqn = 200, 62 and 39 GeV \cite{phenix-jpsi-62-39} together with calculations including suppression and recombination \cite{rapp-jpsi}.}
            \label{fig:phenix-bes}
\end{center}
\end{figure}

PHENIX has also measured the \jpsi \raa at intermediate energies of 62 and 39 GeV (see Fig.~\ref{fig:phenix-bes}) \cite{phenix-jpsi-62-39}. Within the experimental uncertainties, no significant change is observed in $R_{AA}$. 

The almost constant level of suppression observed at mid-rapidity, from the SPS energy of \sqn = 17.3 GeV up to the top RHIC energy of 
$\sqrt{s_{_{NN}}}$~=~200~GeV, can be explained by the interplay between direct \jpsi suppression and coalescence or recombination of $c$ and $\overline{c}$ quarks. As the collision energy increases 
the direct suppression due to color screening in the QGP increases. But at the same time, over this energy range, the $c\overline{c}$ production cross section increases by almost two orders of magnitude increasing the probability of charmonium production by recombination of $c$ and $\overline{c}$ quarks. 
In some models, like the statistical hadronization model, recombination takes place at the hadronization stage \cite{pbm-hsm}. In others, like in the rate equation approach, recombination occurs continuously through the entire evolution of the collision \cite{rapp-jpsi}. 
As an example, calculations based on the rate equation 
approach are shown in Fig.~\ref{fig:phenix-bes}.   Accidentally, the larger direct suppression almost compensates the larger recombination rate yielding an almost constant \raa over this broad energy range, in reasonable agreement with the mid-rapidity PHENIX data.
However, this model has difficulties in reproducing the stronger suppression observed at forward rapidity. In order to do that, it requires a smaller open charm production and a stronger cold nuclear matter suppression  \cite{rapp-jpsi}. 

If a significant fraction of \jpsi is formed by recombination of charm quarks, the \jpsi should inherit  elliptic flow,  $v_2$,  from the charm quarks.  The measurement of \jpsi flow thus provides an independent  and additional support to the recombination scenario. 
At RHIC, the \jpsi $v_2$ measured by STAR in \auau collisions was found to be consistent with zero \cite{star-jpsi-v2}.  But one should note that the predicted $v_2 $ is very small (of the order of 2-3\%) \cite{v2-flow-calculations}  and data of much higher precision are needed before a definite statement can be made. At the LHC, where recombination is expected to play a larger role, the predicted values are somewhat larger  (of the order of 5\%) and the ALICE experiment recently reported a non-zero $v_2$ of \jpsi in semi-central \pbpb collisions which is consistent with calculations \cite{alice-jpsi-v2}.  

The PHENIX experiment has studied the \jpsi production dependence on the system size by varying the colliding nuclei. In \cucu collisions the \jpsi \raa shows a similar behavior to that of \auau collisions when both are compared at the same number of participating nucleons \npart \cite{phenix-cucu}. A real benefit in the study of light systems is that they provide higher precision in the determination of \npart for \npart $\leq$ 100.
Recently, PHENIX measured \jpsi production in Cu+Au and U+U collisions at \sqnr \cite{takao-qm12}. In these systems \jpsi production is tested under different initial geometries and  thus gives additional constraints to theoretical models.

\begin{figure}
\begin{center}
         \includegraphics[width=90mm]{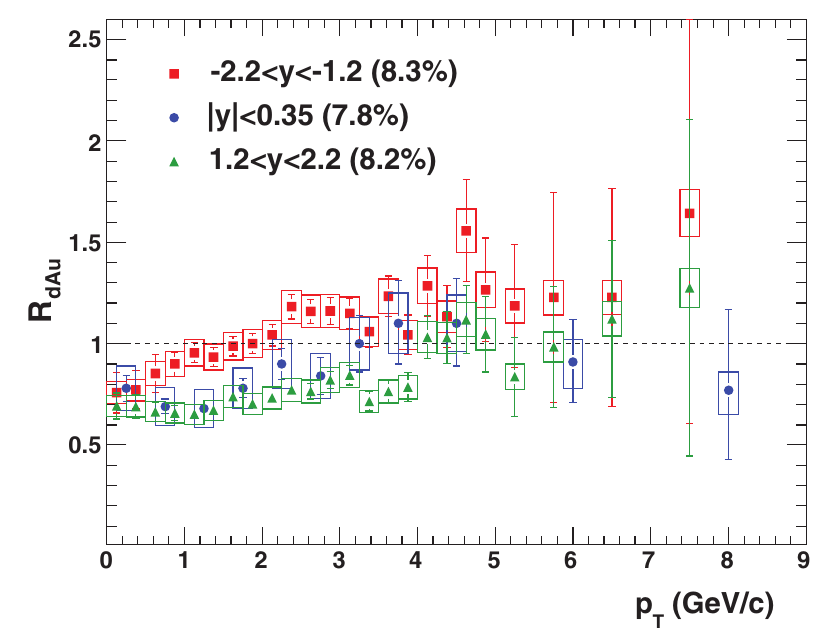}
\caption{Centrality integrated \jpsi $R_{dAu}$ vs \pt measured by PHENIX in \dau collisions at \sqnr \cite{phenix-jpsi-dau}.} 
\label{fig:phenix-cnm}
\end{center}
\vspace{-4mm}
\end{figure}
The observed \jpsi yield can be affected by a variety of cold nuclear matter (CNM) effects including nuclear modifications of the gluon distribution function (shadowing and anti-shadowing), gluon saturation, intial state parton scattering and nuclear absorption (for a review see \cite{cnm-review}). The measurement of \jpsi in a small size system such as \dau is considered the most appropriate way to quantitatively study these CNM effects\footnote{This working hypothesis might be challenged by 
recent results suggesting that effects of hydrodynamic origin occur in \dau and \ppb collisions \cite{phenix-ppg152} $-$ \cite{cms-ppb}.}.
PHENIX has measured \jpsi production in \dau collisions at \sqnr in three rapididty intervals: mid-rapidity ($|y| < $ 0.35), forward rapidity (1.2 $< y <$ 2.2) and backward rapidty (-2.2 $< y < $ -1.2) \cite{phenix-jpsi-dau}. The centrality integrated $R_{dAu}$ vs. \pt  is shown in Fig.~\ref{fig:phenix-cnm} for the three rapidity intervals overlaid. 
A small but significant suppression is seen in the three cases. Whereas at backward rapidity (the Au-going direction) the suppression is seen at low  \pt < 2 GeV/c, the mid- and  forward rapidity intervals show a remarkably similar behavior with a suppression extending up to \pt $\sim$ 4 GeV/c. For all three intervals, \raa is consistent with 1 at \pt $>$ 4 GeV/c, suggesting no sizable CNM effects at high $p_T$. Model calculations including a mixture of CNM effects, such as shadowing, nuclear absorption and Cronin effect, have difficulties in
reproducing all the \dau results \cite{phenix-jpsi-dau}. On the other hand,  recent calculations including only parton \pt broadening and energy loss in the nuclear medium show remarkable agreement with the \pt and centrality dependence of the \jpsi $R_{dAu}$ data at the three rapidity intervals \cite{arleo-jpsi-dau}.

\section{LHC results}
\label{lhc}
\begin{figure}
\begin{center}
         \includegraphics[width=90mm]{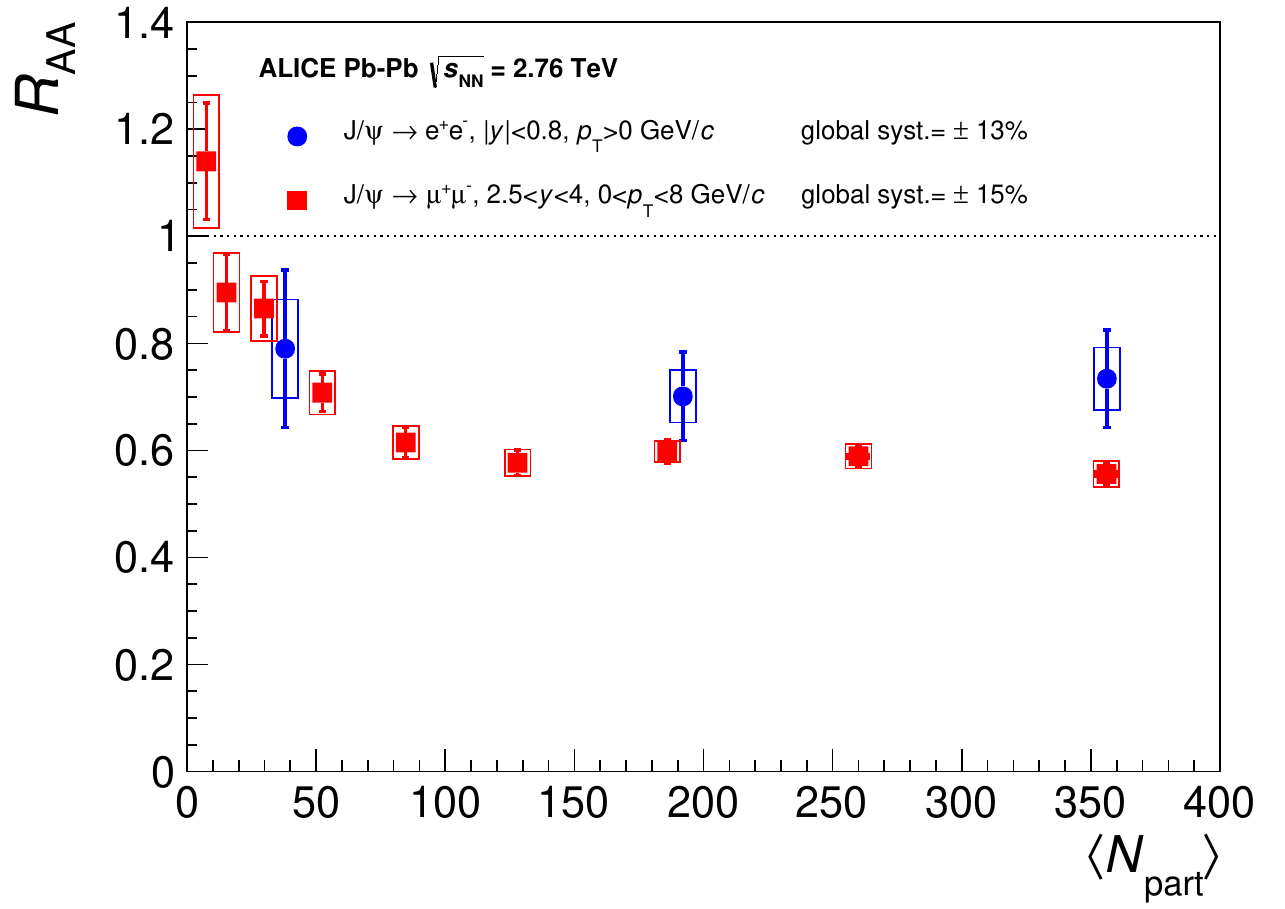}
\caption{\jpsi \raa vs \npart measured by ALICE in \pbpb collisions at       $\sqrt{s_{_{NN}}}$~=~2.76~TeV \cite{alice-jpsi, alice-in-press}.} 
\label{fig:jpsi-alice}
\end{center}
\end{figure}

The charmonium results obtained at the LHC are rather different from those obtained at RHIC. Figure~\ref{fig:jpsi-alice} shows the \jpsi \raa vs. \npart measured by ALICE in \pbpb collisions at \sqn = 2.76 TeV \cite{alice-jpsi, alice-in-press}. The large difference between forward and mid-rapidity
 observed at RHIC (shown in  Fig.~\ref{fig:rhic}) does not seem to be present in the ALICE data.   Furthermore, at forward rapidity, the level of suppression in semi-central or central collisions reaches a value of $\sim$0.6, smaller than the suppression of $\sim$0.2 observed at RHIC (cf. Fig.~\ref{fig:rhic}). 

\begin{figure}[h!]
     \begin{center}
           \includegraphics*[width=70mm, height=60mm] {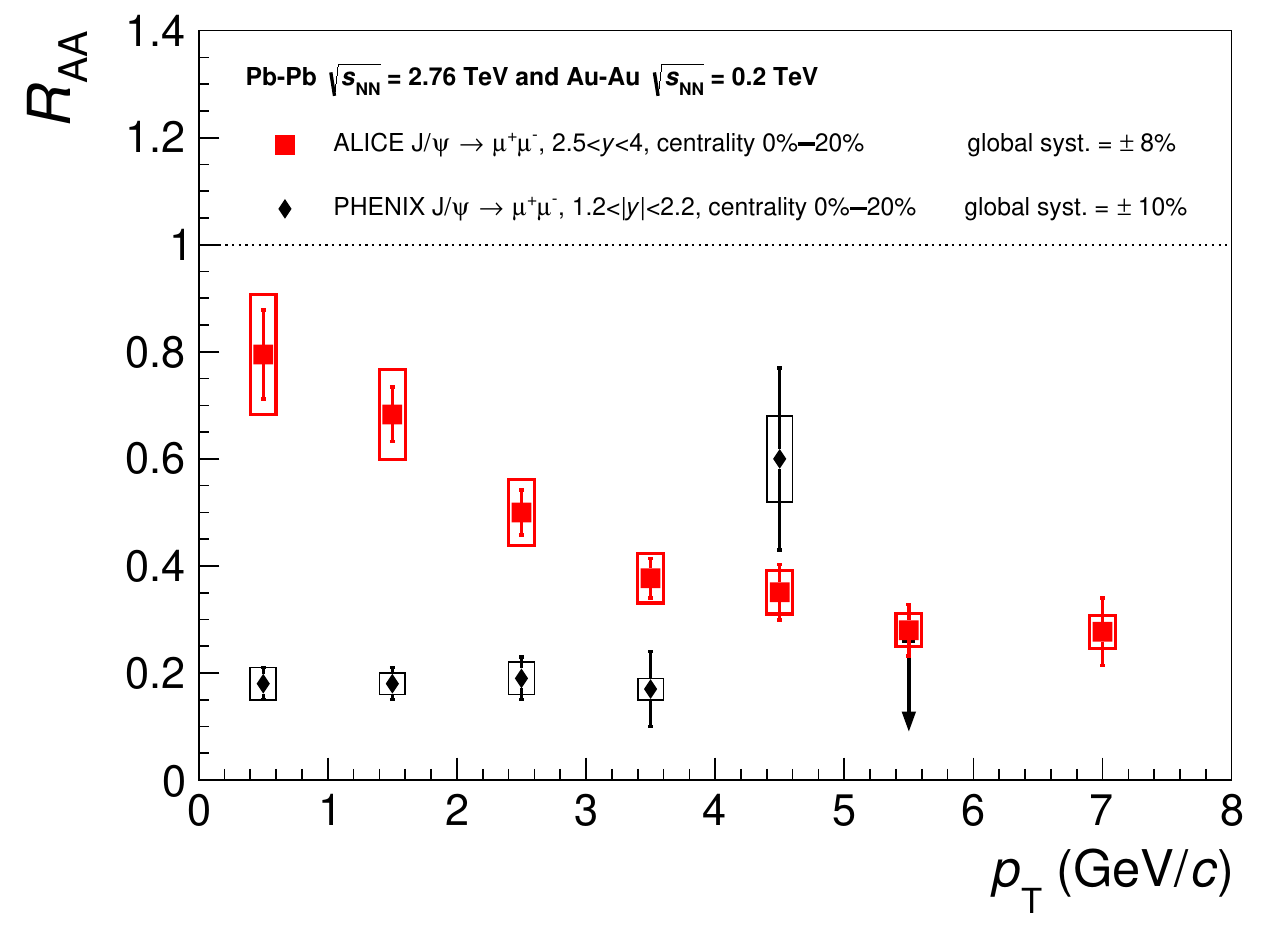}            
           \includegraphics*[width=70mm] {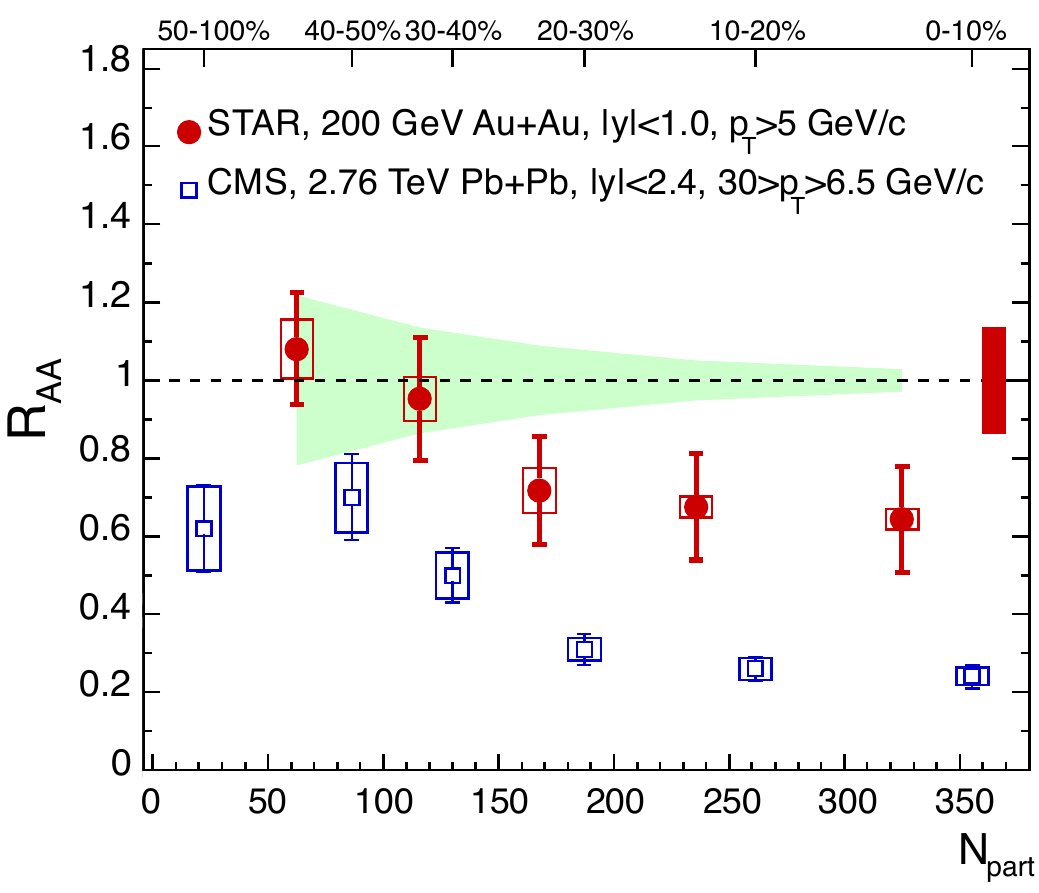}
\vspace{-3mm}
\caption{Left:  \jpsi \raa \pt dependence measured by ALICE at LHC \cite{alice-in-press} and by PHENIX at RHIC \cite{phenix-jpsi-200}.  Right: \raa centrality dependence of high \pt \jpsi measured by CMS \cite{cms-high-pt} and STAR \cite{star-high-pt}.}
           \label{fig:jpsi-pt}
     \end{center}
\vspace{-5mm}
\end{figure}

A large difference is also observed in the \pt dependence of $R_{AA}$. The left panel of Fig.~\ref{fig:jpsi-pt} compares the \raa \pt dependence measured by ALICE at the LHC \cite{alice-in-press} and by PHENIX at RHIC \cite{phenix-jpsi-200} at similar rapidities in central \pbpb and  \auau collisions, respectively. 
At low \pt (\pt $\lesssim$ 4 GeV/c), a stronger suppression is observed at RHIC than at LHC, probably reflecting the larger recombination contribution at the LHC. At high $p_T$, the opposite might be true although the \pt reach of the PHENIX data is not sufficient for a definite statement. 
The \jpsi \raa at high \pt is of particular interest because it might be more sensitive to color screening effects (CNM effects are measured to be low at high \pt as shown in Fig.~\ref{fig:phenix-cnm} and also recombination is expected to be low at high \pt \cite{rapp-jpsi}). 
The right panel of Fig.~\ref{fig:jpsi-pt} compares the \raa centrality dependence for high \pt  \jpsi measured by CMS \cite{cms-high-pt} and STAR \cite{star-high-pt}. A stronger suppression is seen at LHC than at RHIC. However, one should note that the CMS data refer to prompt \jpsi whereas the STAR data are for inclusive \jpsi and thus this comparison might be affected by the \raa of the B mesons feed-down contribution to the \jpsi yield from STAR.
 
CMS pioneered the study of bottomonium states  at LHC \cite{cms-upsilon} as an additional probe to unveil color screening effects in the QGP\footnote{STAR has preliminary results on the production of the non-resolved $\Upsilon$ states (1S + 2S +3S) in \auau collisions at \sqnr \cite{star-upsilon}.}.
\begin{figure}[h]
    \begin{center}
\vspace{-2mm}
        \includegraphics[width=7cm, height=60mm]{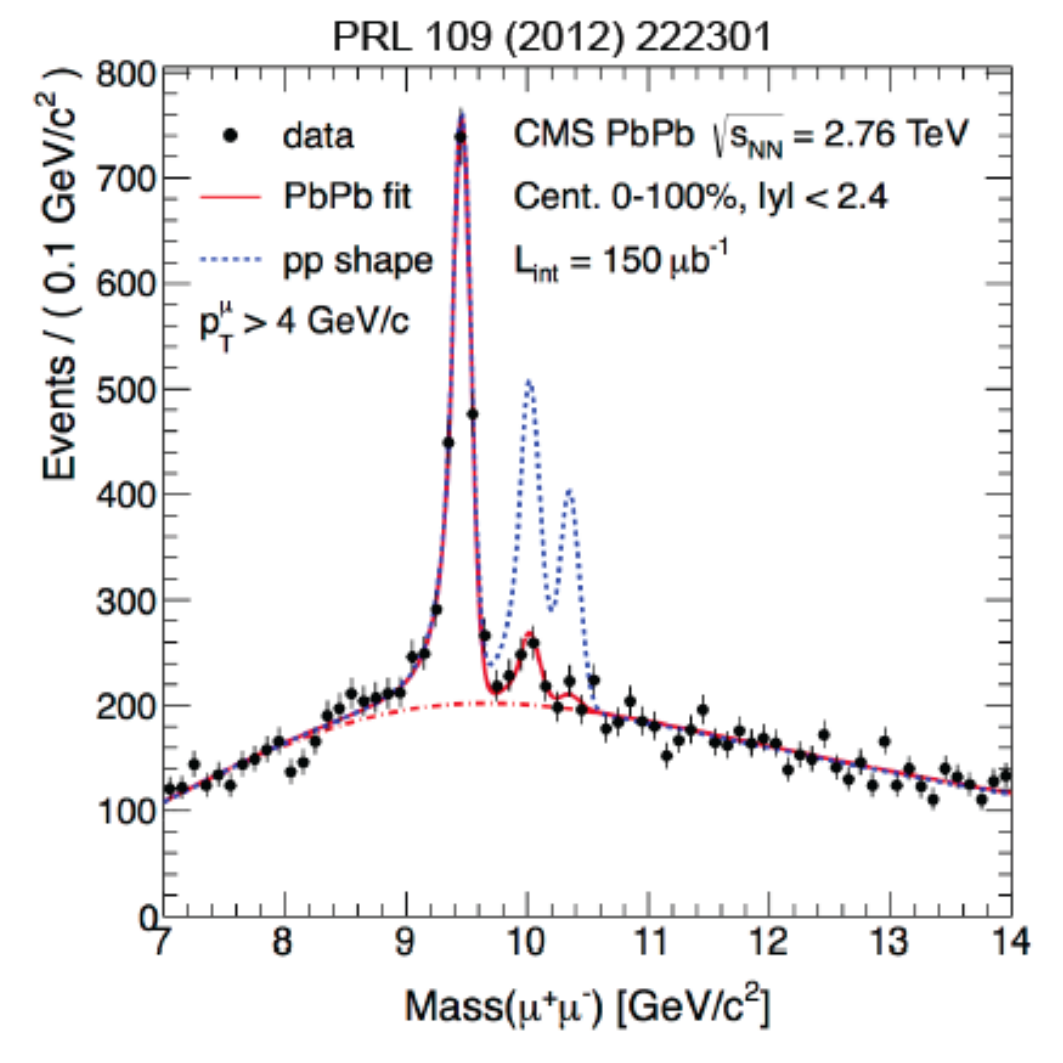}
        \includegraphics[width=7cm, height=60mm]{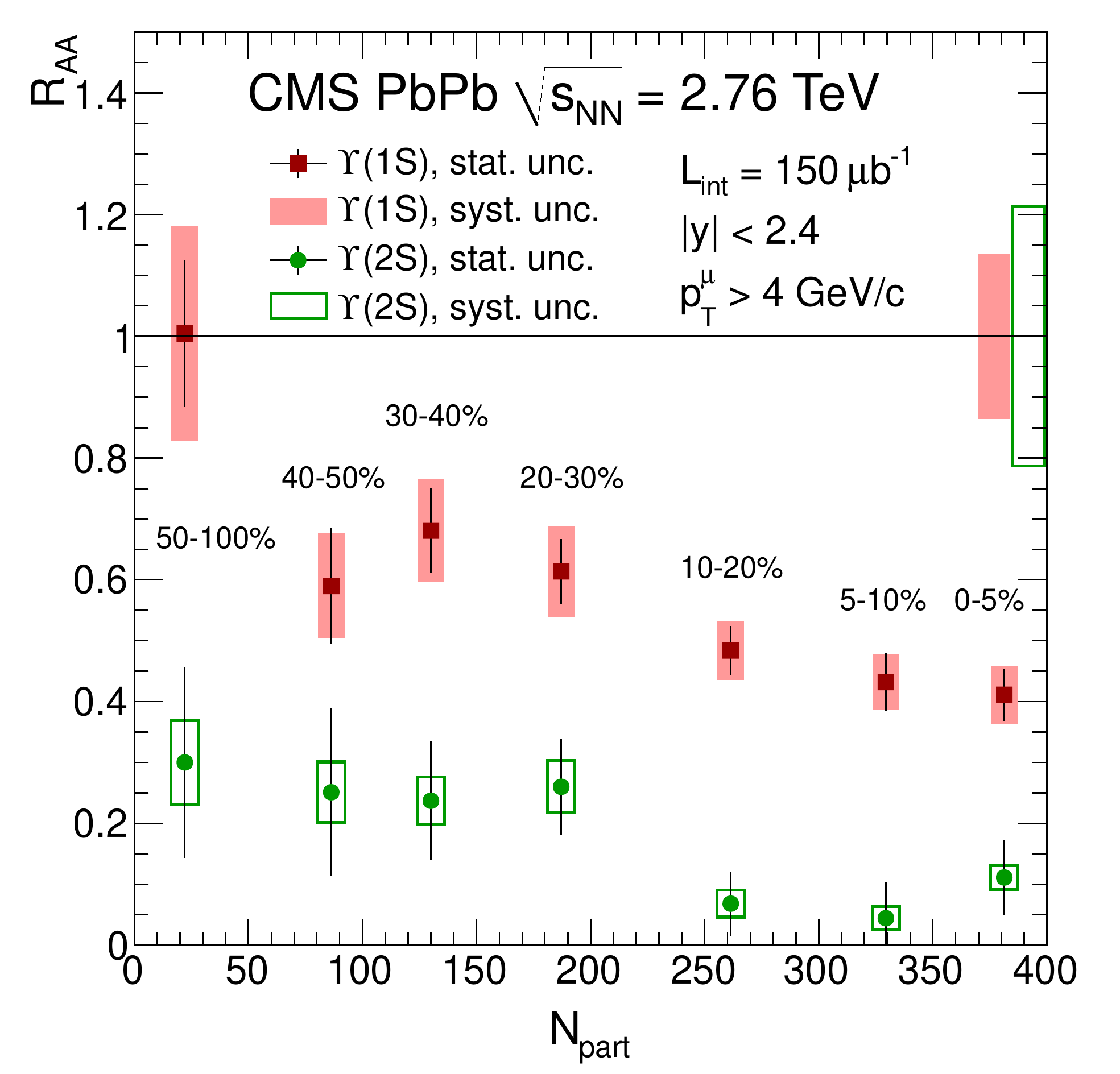}     
   \end{center}
\vspace{-0.8cm} 
       \caption{Left: $\Upsilon$ states measured by CMS in minimum bias \pbpb collisions at \sqn = 2.76 TeV. The p+p mass spectrum shape, normalized to the $\Upsilon$(1s) peak, is also shown \cite{cms-upsilon}. } 
\vspace{-0.1cm}
\label{fig:upsilon}
\end{figure}
The left panel of Fig.~\ref{fig:upsilon} shows the invariant mass spectrum of \mumu pairs in the $\Upsilon$ mass region in minimum bias \pbpb collisions at \sqn = 2.76 TeV \cite{cms-upsilon}. The figure also shows the mass spectrum shape measured in p+p collisions at the same energy  normalized to the peak of the $\Upsilon$(1s) state. The p+p spectrum shows clear separation of the three states  $\Upsilon$(1S, 2S, 3S) demonstrating the excellent mass resolution of the CMS detector. The comparison of the p+p and \pbpb spectra already reveals a clear suppression pattern with the $\Upsilon$(2S) state strongly suppressed and  the $\Upsilon$(3S) state hardly visible.  
The suppression pattern is shown in a quantitative manner in the right panel of Fig.~\ref{fig:upsilon} that displays the $\Upsilon$ \raa centrality dependence. 
$\Upsilon$(1S) is suppressed by a factor of $\sim$2 in central collisions. This is consistent with the assumption that the feed-down states (that account for $\sim$50\% of the $\Upsilon$(1S) yield \cite{upsilon-lhcb}) are fully suppressed. $\Upsilon$(2S) shows a stronger suppression. The $\Upsilon$(3S) state is so suppressed that 
only an upper limit of \raa = 0.10 (with a confidence level of 95\%) is reported for minimum bias \pbpb collisions. This ordering follows the expected sequential melting of the resonances as their binding energy increases,  with the lowest binding energy state, $\Upsilon$(3S), melting first. This  appealing interpretation
needs to be reconsidered after measuring possible nuclear effects in \ppb collisions. It will also be interesting to compare to similar data of resolved $\Upsilon$ states at RHIC energies.

\section{Conclusions}
After more than 25 years of intensive experimental and theoretical effort, the quarkonium saga is still evolving, producing  exciting and surprising results but not fully understood at the quantitative level.
The ongoing systematic study of quarkonia states over a broad energy range and using several collision systems shall ultimately  allow  disentangling the melting of the resonances in the QGP from recombination, cold nuclear matter and other competing effects.

\end{document}